# Room-temperature quantum spin Hall edge state in a higher-order topological insulator

**Authors:** Nana Shumiya[1]*, Md Shafayat Hossain[1]*†, Jia-Xin Yin[1]†, Zhiwei Wang[2,3]*, Maksim Litskevich[1]*, Chiho Yoon[4,5]*, Yongkai Li[2,3], Ying Yang[2,3], Yu-Xiao Jiang[1], Guangming Cheng[6], Yen-Chuan Lin[7], Qi Zhang[1], Zi-Jia Cheng[1], Tyler A. Cochran[1], Daniel Multer[1], Xian P. Yang[1], Brian Casas[8], Tay-Rong Chang[9,10,11], Titus Neupert[12], Zhujun Yuan[13,14,15], Shuang Jia[13,14,15], Hsin Lin[16], Nan Yao[6], Luis Balicas[8], Fan Zhang[4], Yugui Yao[2,3], M. Zahid Hasan[1,17]†

**Affiliations:**

[1]Laboratory for Topological Quantum Matter and Advanced Spectroscopy (B7), Department of Physics, Princeton University, Princeton, New Jersey, USA.

[2]Centre for Quantum Physics, Key Laboratory of Advanced Optoelectronic Quantum Architecture and Measurement (MOE), School of Physics, Beijing Institute of Technology, Beijing 100081, China.

[3]Beijing National Laboratory for Condensed Matter Physics and Institute of Physics, Chinese Academy of Sciences, Beijing 100190, China.

[4]Department of Physics, University of Texas at Dallas, Richardson, Texas 75080, USA.

[5]Department of Physics and Astronomy, Seoul National University, Seoul 08826, Korea.

[6]Princeton Institute for Science and Technology of Materials, Princeton University, Princeton, NJ, USA.

[7]Department of Physics, National Taiwan University, Taipei, 10607, Taiwan.

[8]National High Magnetic Field Laboratory, Tallahassee, Florida 32310, USA.

[9]Department of Physics, National Cheng Kung University, Tainan, Taiwan.

[10]Center for Quantum Frontiers of Research and Technology (QFort), Tainan, Taiwan.

[11]Physics Division, National Center for Theoretical Sciences, Hsinchu, Taiwan.

[12]Department of Physics, University of Zurich, Winterthurerstrasse, Zurich, Switzerland.

[13]International Center for Quantum Materials, School of Physics, Peking University, Beijing, China.

[14]CAS Center for Excellence in Topological Quantum Computation, University of Chinese Academy of Sciences, Beijing, China.

[15]Beijing Academy of Quantum Information Sciences, Beijing, China.

[16]Institute of Physics, Academia Sinica, Taipei 11529, Taiwan.

[17]Lawrence Berkeley National Laboratory, Berkeley, California 94720, USA.

†Corresponding authors, E-mail: mdsh@princeton.edu; jiaxiny@princeton.edu; mzhasan@princeton.edu.

*These authors contributed equally to this work.

**Room-temperature realization of macroscopic quantum phase is one of the major pursuits in fundamental physics[1,2]. The quantum spin Hall phase[3-6], a topological quantum phase that features a two-dimensional insulating bulk and a helical edge state, has not yet been realized at room temperature. Here, we use scanning tunneling microscopy to visualize a quantum spin Hall edge state on the surface of the higher-order topological insulator $Bi_4Br_4$. We find that the atomically resolved lattice exhibits a large insulating gap of over 200meV, and an atomically sharp monolayer step edge hosts a striking in-gap gapless state, suggesting the topological bulk-boundary correspondence. An external magnetic field can gap the edge state, consistent with the time-reversal symmetry protection inherent to the underlying topology. We further identify the geometrical hybridization of such edge states, which not only attests to the $Z_2$ topology of the quantum spin Hall state but also visualizes the building blocks of the higher-order topological insulator phase. Remarkably, both the insulating gap and topological edge state are observed to persist up to 300K. Our results point to the realization of the room-temperature quantum spin Hall edge state in a higher-order topological insulator and encourage further exploration of high-temperature transport quantization of the topological phase reported here.**

A topological insulator is a material that behaves as an insulator in its interior but whose surface contains protected conducting states[3, 7-11]. A two-dimensional topological insulator features time-reversal symmetry-protected helical edge states residing in an insulating bulk gap (see Fig. 1**a** for both the real-space and momentum-space pictures of the helical edge state), and exhibits the quantum spin Hall effect accordingly[4-6, 12-15]. The helical edge state features dissipationless electron channels along the sample edges (Fig. 1**a**), which is of great interest in energy-saving technology and quantum information science. Equipped with high spatial resolution, electronic detection, and magnetic field tunability, the state-of-the-art scanning tunneling microscopy has been very powerful to discover and elucidate topological edge states in quantum materials[15-28]. Among topological insulator candidates, $Bi_4Br_4$ has a layered structure with van der Waals like bonding, and has been proposed to feature a large insulating gap and weak interlayer coupling; thus monolayer $Bi_4Br_4$ has the potential to realize the high-temperature quantum spin Hall state[29-34] in both freestanding[29] and bulk[34] environments. While pioneering work using the angle-resolved photoemission technique[35] set out to resolve the topological boundary mode from crystalline steps, it remains elusive whether the observed boundary mode signal arises from the side surfaces or atomic step edges of the crystal. The magnetic field response and temperature robustness of the predicted quantum spin Hall state are also largely unexplored, which will provide indispensable information for the underlying quantum topology and future applications of this quantum material. Therefore, a real-space experimental investigation of the nature of the edge state with atomic layer spatial resolution, magnetic field tunability, and temperature control is highly desirable. In this work, we use scanning tunneling microscopy to demonstrate and elucidate its insulating gap and topological edge state.

The crystal structure of $\alpha$-Bi$_4$Br$_4$ is shown in Fig. 1**b**, which exhibits an interlayer stacking along the *c*-axis and quasi-one-dimensional chains running along the *b*-axis within each layer. Notably, two adjacent layers tilt towards different directions, as evidenced by our scanning transmission electron microscopy image visualizing the AB stacking as shown in Fig. 1**c**. For our scanning tunneling microscopy measurements, we cleave the single-crystal *in-situ* and measure at a temperature of 4.2K, obtaining atomically clean surfaces as shown in Fig. 1**d**. Clearly, the topographic image breaks mirror symmetry along the *a*-axis, and its symmetry-breaking direction allows us to determine this layer as the A layer as defined in Fig. 1**c**. The determination of the mirror symmetry breaking of the lattice is crucial for us to perform an analysis of the step edge geometry. The tunneling differential conductance (d$I$/d$V$) that measures the local density of states directly reveals a large insulating gap of $\Delta$ = 260 meV on the atomic lattice (Fig. 1**e**). Such a large gap is consistent with the bulk insulating gap calculated from first-principles[32] and observed by angle-resolved photoemission spectroscopy[35]. At higher energies, the

To further explore the topological nature of the insulating gap, we check the topological bulk-boundary correspondence in Fig. 2**a**. We perform d$I$/d$V$ spectroscopic maps at a monolayer step edge identified by a topographic image (Fig. 2**a**). As the electronic structure at the Fermi level is often regarded to be most important for a quantum material, we focus on presenting the real-space differential conductance maps at the Fermi level. We find that the step edge exhibits pronounced edge states within the insulating gap (Fig. 2**b** left panel for the spectroscopic imaging and Fig. 2**d** for the tunnelling spectra). The left panel of Fig. 2**b** shows the decay of the edge state on the crystal side of the step edge. While the edge state decays more sharply on the vacuum side, it decays with a characteristic length of $r_0 \approx 2.1$nm on the crystal side. The penetration depth of a topological boundary mode can be generically estimated[8] by $r_0 = 2\hbar v_F/\Delta$, where $v_F$ is the bulk avoided Fermi velocity. This allows us to further estimate the Fermi velocity to be 0.27eV•nm, which is of the same order of magnitude as the one calculated by first-principles calculations[32] and determined by angle-resolved photoemission experiment[35]. The d$I$/d$V$ spectrum taken at the step edge clearly shows the existence of the gapless in-gap state (Fig. 2**d**). At the Fermi energy, the in-gap state exhibits a dip feature. Such a zero-bias anomaly has also been consistently reported in other quantum spin Hall edge state systems[13, 19-27, 36], as a result of the Tomonaga-Luttinger liquid behavior due to edge electron-electron interactions.

Furthermore, the quantum topology of the system has been proposed to be protected by the time-reversal symmetry[29, 34], which can be broken by applying an external magnetic field. When we apply a strong magnetic field perpendicular to the *ab* plane, we find that the in-gap state measured at the edge is suppressed substantially (Fig. 2**c** for $B$ = 4T and $B$ = 8T), and an energy gap progressively develops for the edge state (Figs. 2**e** and **f**). The magnetic field essentially opens a gap at the edge state Dirac point. The energy gap position in Fig. 2**f** suggests that the Dirac point is slightly above the Fermi level. From our field-dependent data, we estimate the gap opening rate of 3meV/T, which amounts to a large *c*-axis Landé g-factor of 53. We note that for this material, the Dirac point energy and band dispersion of the edge state may be subject to band bending induced by the probe tip[37,38]. The bulk-boundary correspondence together with the response to the breaking of time-reversal symmetry demonstrated here, provide experimental evidence for the topological nature of the system[29].

To further demonstrate the $Z_2$ topological nature of the monolayer step edge, we study their quantum hybridization. The hybridization of two $Z_2$ edge modes is destructive and opens an energy gap. Specific to this material, rooted in the fact that the inversion center is in the monolayer instead of in the center of a bilayer[34], the left and right AB-bilayer edges have different geometries (Fig. 3**a**). On the left, the edges of the A and B layers have a facing angle larger than 180°, while on the right the facing angle is smaller than 180°. This inversion asymmetry leads to stronger hybridization of the monolayer edge states on the right[34]. Similarly, for a BA-bilayer, it can be inferred that the left side has stronger hybridization (not shown). Notably, this interlayer edge state hybridization has been pointed out as a key building block of the bulk higher-order topological insulator[39-42] that exhibits helical hinge states[34] (Fig. 3**b**). Figure 3**c** shows a case region containing both a monolayer step edge and a left edge of AB-bilayer. The associated dI/dV mapping confirms the existence of in-gap edge states in both cases. To visualize the asymmetric hybridization of the edge states depicted in Fig. 3**a**, we extensively scan the crystal to find an area containing both left and right AB-bilayer step edges as shown in Fig. 3**d**. The differential spectra taken at the two AB-bilayer step edges, as shown in Fig. 3**e**, reveal their dramatically different behaviors. The in-gap state for the right-AB bilayer edge is substantially suppressed compared with that on the left-AB bilayer edge. The inversion asymmetry of the bilayer edge state is further visualized in Fig. 3**f**, which demonstrates the dI/dV maps of the same area as in Fig. 3**d**, taken at two representative energies within the insulating gap. It is shown that the edge state mainly shows up for the left-AB bilayer edge, in agreement with the picture in Fig. 3**a**. The geometrical dependence of bilayer step edge states not only attests to the $Z_2$ topology of the quantum spin Hall state within each monolayer but also demonstrates the building block of the bulk higher-order topological insulator.

Having observed the quantum spin Hall edge state at low temperature, we further explore the edge state by progressively increasing the temperature. In Fig. 4**a** we show dI/dV spectra of the lattice and monolayer step edge, taken at high temperatures, 100 K, 200 K, and 300 K. Similar to our d*I*/d*V* data at $T = 4.2$ K, at all three temperatures, both the insulating gap and the in-gap edge state persist. Although, the insulating gap appears to shrink to ~230 meV (by ~10%) owing to thermal broadening and other factors, the Fermi level remains in the gap. The temperature robustness of the topological edge state is further demonstrated in Fig. 4**b** where we show topographic images around a step edge and d*I*/d*V* maps of the same area, taken at zero energy for the different temperatures. In all cases, the step edge exhibits a pronounced localized state. The presence of the topological edge state at room temperature is further substantiated in the d*I*/d*V* spectra (Fig. 4**c**) taken along a line perpendicular to the edge shown in the top panel of Fig. 4**b**. At and very near the position of the edge, the in-gap edge states emerge, whereas on the surface away from the edge the spectra are gapped near the Fermi energy. Our observation of the quantum spin Hall edge state and the insulating gap at the Fermi level at room temperature is unprecedented, and our work microscopically demonstrates the building block of a higher-order topological insulator. In magnetically doped topological insulators, the topological gap is spatially inhomogeneous, which severely limits the lifetime of the chiral edge states[43]. In striking contrast, the large bulk insulating gap in this stoichiometric material is shown to be rather homogeneous in our study, and the lifetime of helical edge states was recently reported to be ultralong[44]. The room-temperature quantum spin Hall state observed here does not require extreme conditions, including giant pressure[2] or high-magnetic field[1]; therefore, it holds great potential for next-generation quantum technologies.

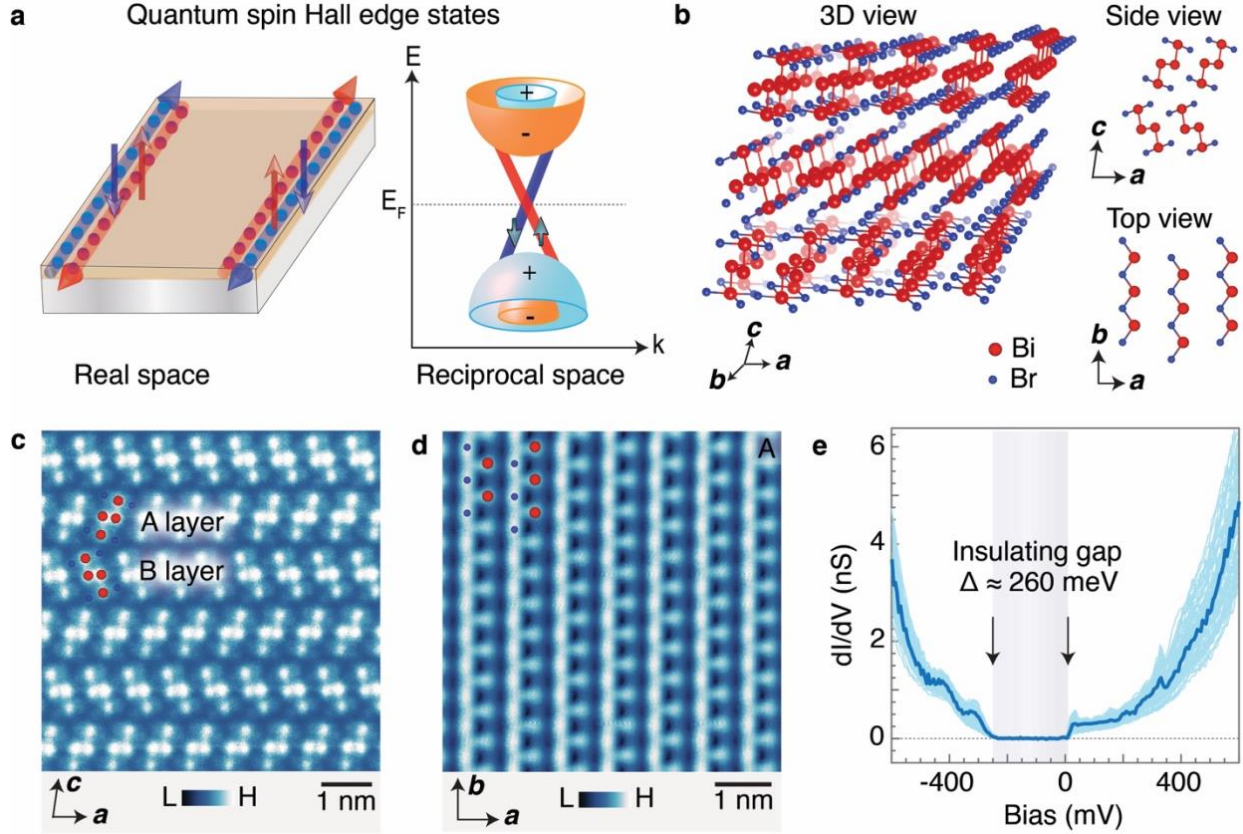

**Figure 1. Observation of a large insulating energy gap. a,** Schematic of the quantum spin Hall edge state. The top panel illustrates the counter-propagating helical edge states with spin up (red) and spin down (blue) dissipationless channels in the real space. The bottom image shows the same topological edge states with red and blue bands projected in the surface momentum space. The edge states originate from the bulk band inversion, which is sketched by the orange and light-blue bands. '+' and '-' denote the bulk bands with even and odd parities, respectively. **b,** Three-dimensional crystal structure, top view of monolayer (bottom right) and side view of the bulk (top right) $\alpha$-$Bi_4Br_4$. **c,** Scanning transmission electron microscope image from a lateral perspective, showing the atomic interlayer AB stacking. **d,** Atomically resolved scanning tunneling microscopy image of the cleaving surface showing the type-A lattice. **e,** Tunneling spectroscopy taken at 4.2 K, revealing an insulating gap. Light blue curves are the differential spectra taken at different positions on the surface; the dark blue curve denotes the average spectra.

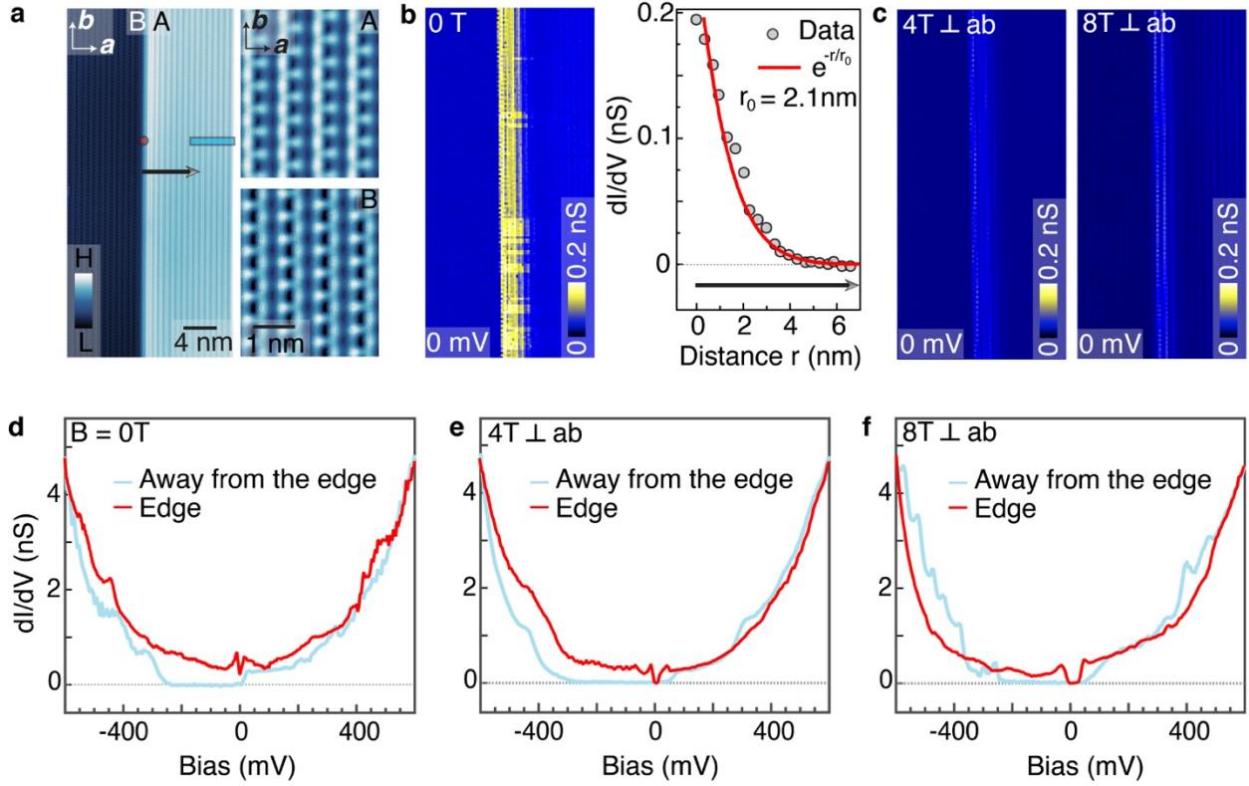

**Figure 2. Observation of quantum spin Hall edge state. a**, Topographic image of an atomically sharp monolayer AB step edge. An enlarged atomically resolved view of A and B surfaces are shown on the right panels. **b**, Left panel d$I$/d$V$ map taken at $B = 0$ T (V = 0 mV) showing a pronounced in-gap edge state. Right panel shows the intensity distribution of differential conductance taken at 0 mV away from the step edge (corresponding location is marked on the topographic image in panel **a** with a black line; the direction of the scan is marked with an arrow). The red curve shows the exponential fitting of the decay of the state away from the edge. **c**, d$I$/d$V$ maps taken at $B = 4$ T and 8 T (V = 0 mV) shown using the same color scale as in panel **b**. **d**, **e**, and **f**, Differential spectra at $B = 0$ T, 4 T, and $B = 8$ T, respectively, further highlighting the effect of time-reversal symmetry breaking on the edge state. Red and blue curves denote the spectra taken at the edge and far away from the edge (locations marked in the topography image in panel **a**; blue curves are averaged over the marked region), respectively. Spectra at different magnetic fields are taken at the same locations, and using standard lock-in technique with a frequency of 977 Hz and a junction set-up of $V = -600$ mV and $I = 0.5$ nA, and a root mean square oscillation voltage of 1mV. At $B = 4$ T and 8 T, the edge state is clearly suppressed, and a gap gradually develops at the Fermi level as a function of the magnetic field.

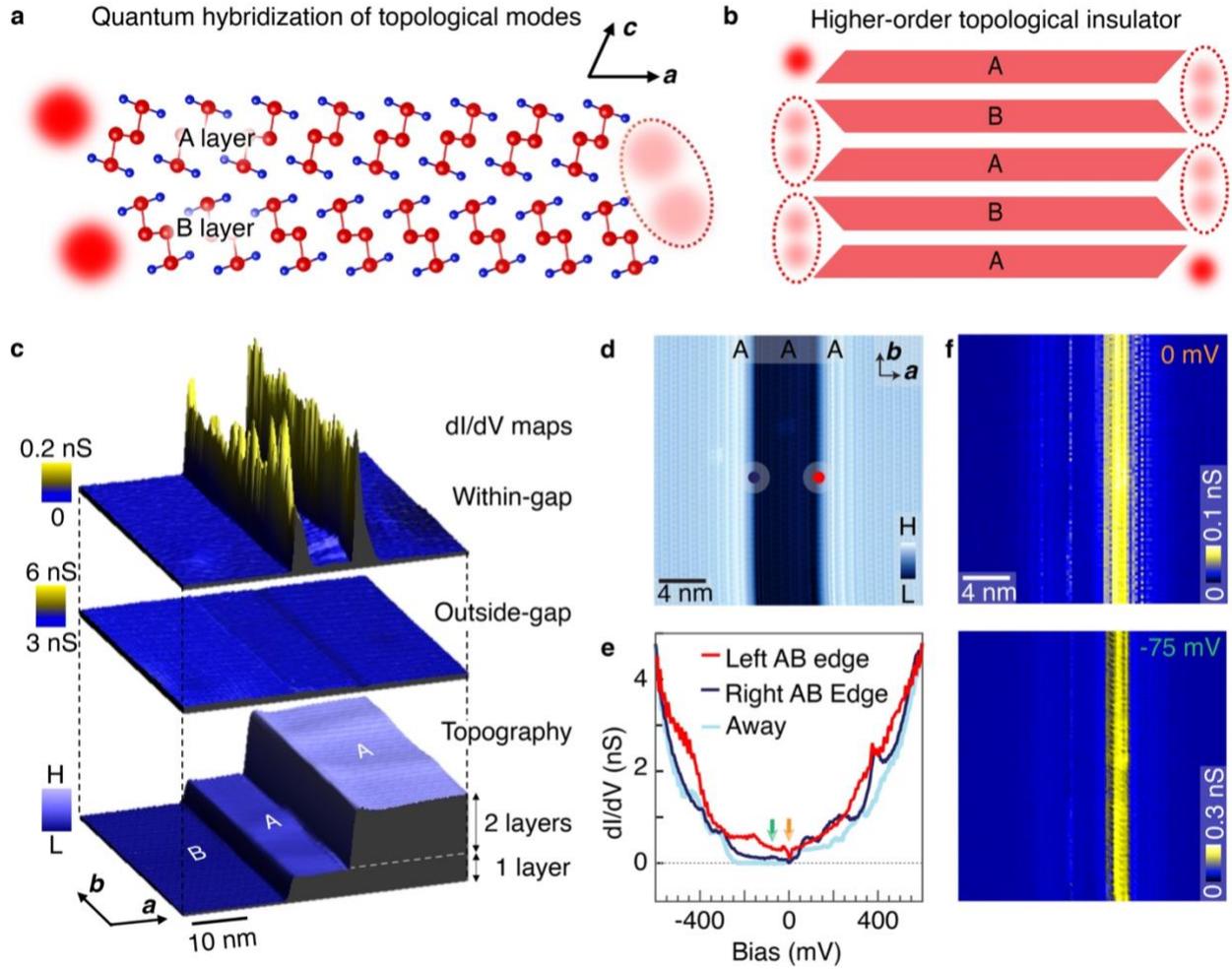

**Figure 3. Geometrical hybridization of quantum spin Hall edge states. a**, Side view of an AB-bilayer. On the left, the edges of A and B layers have a facing angle larger than 180°, while on the right, the facing angle is smaller than 180°. This geometrical difference leads to stronger hybridization of the two monolayer edge states (red spheres) on the right. The hybridization of quantum spin Hall edge state is destructive, as illustrated by the lighter color of the edge states (red spheres). **b,** Construction of a higher-order topological insulator based on the hybridization mechanism in **a**. The higher-order topological insulator in this case exhibits two hinge modes on the top left and bottom right corners. **c,** Spectroscopic imaging of the crystalline steps. The bottom panel shows the topographic image of two-step edges. The middle panel shows the corresponding differential conductance map obtained at energy outside the insulating gap. The top layer shows the corresponding differential conductance map obtained at energy inside the insulating gap, revealing pronounced edge states. **d**, Topographic image of two adjacent bilayer atomic step edges forming a trench. **e**, Differential spectra taken at the left AB bilayer step edge (red), right AB bilayer step edge (violet), and away from the edges (blue) reveal striking differences between the two step edges. Red and violet dots in panel **d** denote the respective positions on the left and right AB edges where the differential spectra are taken. The left AB edge exhibits a pronounced in-gap state, whereas on the right AB edge, the density of states at the bulk gap is largely suppressed. **f**, Differential conductance map, taken at $V = 0$ mV and -75 mV (marked with color-coded arrows in the differential spectra in panel **e**) shown in two layers. $V$ = 0 mV and -75 mV, both of which are inside the insulating gap.

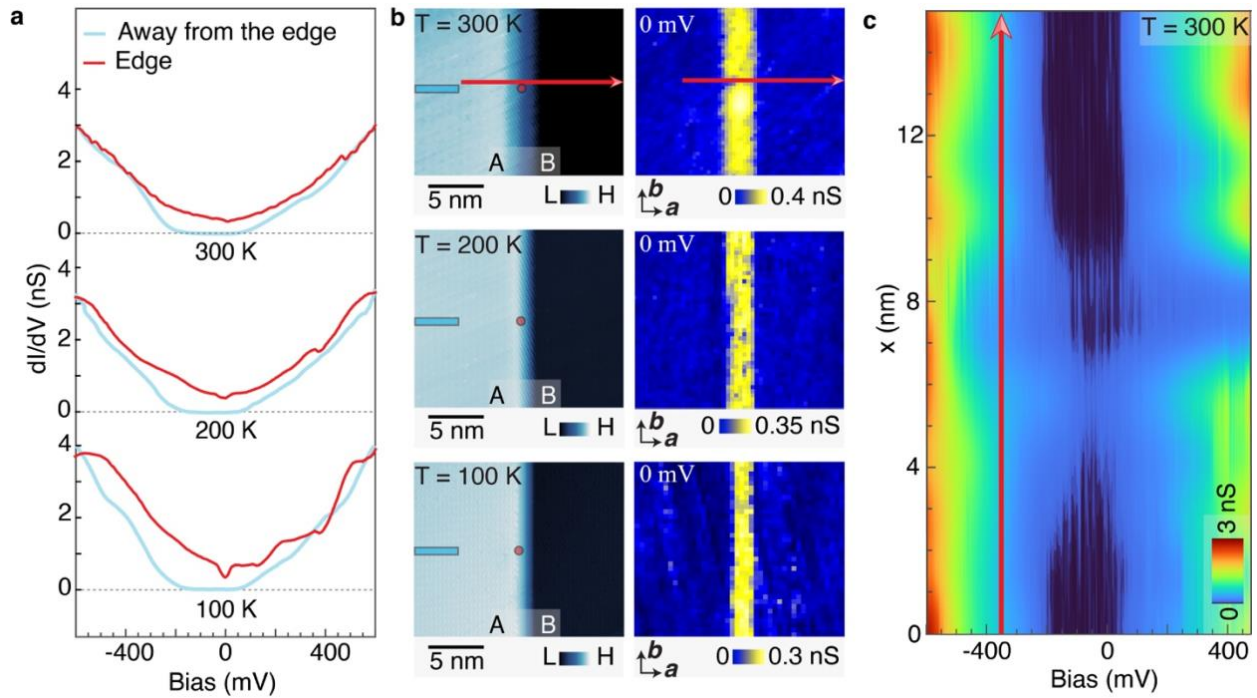

**Figure 4. Room-temperature quantum spin Hall edge state. a**, Temperature-dependent differential spectra taken on surface and edge, denoted with blue and red curves (locations marked in the corresponding topography images in panel **b**; blue curves are averaged over the blue, marked regions), respectively. Spectra are offset for clarity for different temperatures. **b**, Topography and corresponding differential conductance maps taken on a monolayer step edge at $T = 300$ K, 200 K, and 100 K (V = 0 mV), capturing the temperature-robustness of edge states. **c**, Intensity plot of a series of line spectra at $T = 300$ K, taken along the *a*-axis direction (marked on the corresponding topographic image in panel **b** with a red line; the direction of the scan is marked with an arrow), reveals the presence of gapless edge state at room temperature.

## Methods
### Crystal growth
High-quality single crystals of α-Bi$_4$Br$_4$ were grown by a self-flux method. High-purity BiBr$_3$ powder (4N) and Bi shot (6N) were mixed in a molar ratio of 1:16 and then sealed in an evacuated quartz tube. The tube was put into a muffle furnace and heated to 500°C slowly, at which it was held for 24h. After that, it was cooled to 320°C at a rate of 5°C/h and followed by to 275°C at a rate of 2°C/h. Finally, needle-shaped Bi$_4$Br$_4$ single crystals were obtained after the Bi flux was removed via centrifugation. In order to obtain high-quality crystals, we performed surface cleaning procedures to remove the oxide layers formed in air on the raw shots of Bi, i.e., Bi shots were sealed in a quartz tube filled with 0.8 atm hydrogen and then annealed at 240 °C for 12 h.

### Scanning tunnelling microscopy
Single crystals were cleaved mechanically in situ at 77 K in ultra-high vacuum conditions, and then immediately inserted into the microscope head, already at $^4$He base temperature (4.2 K). More than 30 single crystals have been cleaved for this study. For each cleaved crystal, we explored surface areas over 5 μm × 5 μm to search for atomic flat surfaces. Topographic images in this work were taken with the tunnelling junction set-up $V = -600$ mV and $I = 0.05$ nA in the constant current mode. Tunnelling conductance spectra were obtained with a commercial Ir/Pt tip using standard lock-in amplifier techniques with a lock-in frequency of 977 Hz and a junction set-up of $V = -600$ mV and $I = 0.5$ nA, and a root mean square oscillation voltage of 1mV. Tunnelling conductance maps were obtained with a junction set-up of $V = -600$ mV and $I = 0.3$ nA, and a root mean square oscillation voltage of 10 mV. The magnetic field was applied with a zero-field cooling method. For the field-dependent tunnelling conductance map, we first withdrew the tip away from the sample, and then slowly ramped the field to the desired field. Then we reapproached the tip to the sample, found the same atomic area and then performed spectroscopic mapping at this magnetic field. For variable temperature measurement, we first withdrew the tip from the sample, and then raised the temperature and stabilized the temperature for 12 hours, after which we reapproached the tip to the sample to perform tunneling experiments.

**Extended scanning tunnelling microscopy data**

Our Fig. 1**e** shows certain spreading of the tunnelling spectra at high energies. As shown in Extended data Fig. 1, both the d$I$/d$V$ map and spectra show a detectable intra-unit cell variation, particularly at high energies. Similar behavior is observed in other quantum materials whose surfaces are composed of different atoms. We further show extended dI/dV maps in Extended Data Fig. 2 with $V = 0$mV (at Fermi energy), $V = -75$mV (near the middle of the bulk insulating energy gap), and $V = -600$mV (outside the insulating energy gap). For a 1D helical edge state, the magnetic field along its spin direction only shifts the Dirac point momentum, whereas the magnetic field in a perpendicular direction can induce a Zeeman gap[3]. Our 1T in-plane magnetic field data are consistent with this picture, as shown in Extended Fig. 3. We show a high-resolution image of the monolayer step edge in Extended Fig. 4. In comparison with the image far from step edges, we show that there is no detectable change of the inter-stripe distance near the step edge. The step height from the topographic image enhances 5% at the position of the step edge when compared with that away from the step edge. As the topographic image convolutes atomic structure and the integrated density of states, this enhancement is likely due to the contribution from emerging step edge states. In Extended Data Fig. 5, we show tunnelling spectroscopy on both left and right monolayer step edges, and both data reveal gapless edge states. This observation is different from the bilayer step edge, and supports the quantum spin Hall state nature of the monolayer step edge state. Through our systematic search for higher step edge layers, we observe a four-layer step edge as shown in Extended Data Fig. 6. At the edge, we observe gapless tunnelling data, which serves as a candidate gapless hinge state signal. Finally, we show individual d$I$/d$V$ spectrum taken across a monolayer step edge at 300 K in Extended Data Fig. 7.

**Extended theoretical elaboration**

Informed by the bulk crystalline structure of $Bi_4Br_4$, first-principles calculations were performed with the VASP code[45, 46] to obtain the electronic band structures and analyze the topological character of the bands. To obtain more accurate information on band gaps and band inversions, the Heyd-Scuseria-Ernzerhof hybrid functional method[47] was applied. The Wannier90 code[48] was employed to construct the maximally localized Wannier functions for the *p*-orbitals of Bi and Br, which were used to build an *ab initio* tight-binding model[32]. The band structures of monolayer, few-layers, step edges, top surface, side surfaces, and bulk were all calculated through this model.

By using the *ab initio* model, the one-layer-step-edge-projected band structure can be obtained in the bulk environment. In Extended Data Fig. 8, it is clear that the helical edge state along the chain direction (***b***) is present in a 270 meV gap. In bulk, the Dirac velocity in the ***a*** direction is 0.2 eV·nm, as estimated from the low-energy bulk bands, consistent with that evaluated from our experiment as discussed in the main text. The linear edge-state spectrum results in a constant density of states (DOS) inversely proportional to the Fermi velocity. The constant DOS can be probed via the tunneling conductance in the ideal single-particle scenario. However, fluctuations can arise due to the presence of electron-electron interactions, electron-phonon couplings, impurities, lattice defects, etc. In particular, the presence of electron-electron interactions gives rise to Luttinger liquid behavior, which yields a zero-bias dip in the tunneling conductance at Fermi energy.

As a higher-order topological insulator, it has been revealed that $Bi_4Br_4$ has a hinge-state profile depending on how its surfaces are terminated[34]. For a sample with perfectly uniform cuts at the (001), (00-1), (100), and (-100) surfaces, there is always one hinge at the top surface and one hinge at the bottom surface that host a helical hinge state. For an even-layer system, due to the inversion asymmetry, such two special hinges are located on the same side surface, either the (100) or (-100) surface, as exemplified by the twenty-layer case in Extended Data Figs. 9 **a** and **c**. The orange bands are the gapped (100) and (-100) side surface states, whereas the red bands are two helical hinge states localized at the two hinges of the (100) side surface. This feature is even robust in the thinnest limit – the bilayer case, as shown in Extended Data Figs. 9 **b** and **9 c**.

Because the inversion center is in the monolayer, the interlayer couplings are oppositely dimerized at the (100) and (-100) side surfaces. It follows that for a bilayer the two helical edge states of the two constituent monolayers are coupled relatively stronger at one side while relatively weaker at the other side. The stronger coupling produces an edge state gap, whereas the weaker coupling, which turns out negligibly small, has little impact on the gapless edge states. These results for the bilayer case can be confirmed by calculations, as shown in Extended Data Figs. 9 **b** and **c**. In the thick limit, nevertheless, the nearly gapless edge states at one side evolve into the protected gapless hinge states, while the strongly gapped edge states at the other side evolve into the gapped side surface states, as exemplified by the twenty-layer case in Extended Data Figs. 9 **a** and **c**.

Our first-principles calculation in Extended data Fig. 9b suggests that the hybridization gap for the right edge state of the AB bilayer is around 25 meV. This is of the same order as the energy range of the full suppression of the tunneling density of states (DOS) near the Dirac point that is accidentally around the Fermi level. Beyond this energy range, one possible explanation for the reduction of tunneling DOS is the tunneling geometry of the right edge; Fig. 3d shows the topographic image of two adjacent bilayer step edges forming a trench that is close to a "凹" shape. In the real-space picture, the hybridization of the two right edge states produces hybridized orbitals that locate deeper into the 凹" shape. This orbital geometry naturally reduces the tunneling probability to the tip, as compared with a monolayer edge state or the left edge state of an AB bilayer. On the other hand, the tunneling probability from the projected bulk states may be less affected by the edge geometry. Thus, the tunneling signal at the right edge of the AB bilayer, which contains contributions from both the projected bulk states and the edge states, shows a full suppression of DOS in the small energy range of the hybridization gap and a partial reduction of tunneling DOS elsewhere.

Finally, we discuss the magnetic field effect on the monolayer helical edge state. The gapless nature of the helical edge state, i.e., the Kramers degeneracy at the Dirac point, is protected by time-reversal symmetry. When time-reversal symmetry is broken, e.g., by applying a magnetic field, a Zeeman energy gap may be opened at the Dirac point[3]. Consider a simplified model Hamiltonian for the helical edge state: $H_0 = v\hbar k_y \sigma_y$, where $\sigma_i$ ($i = x, y, z$) are the Pauli matrices, $v$ is the Dirac velocity, and $\hbar$ is the Planck constant. The corresponding dispersions are $E_\pm = \pm v\hbar k_y$, and the two bands linearly cross at the Dirac point at $k_y = 0$. When the magnetic field is applied along the $y$-axis, the resulting Zeeman term can be modeled as $H_\parallel = \alpha_y B_y \sigma_y$, where $\alpha_i$ ($i = x, y, z$) are coefficients related to the Landé g-factor. In this case, the dispersions become $E_\pm = \pm v\hbar(k_y + \alpha_y B_y/v\hbar)$, i.e., the band crossing point is shifted along the $k_y$-axis. When the magnetic field is applied along the $z$-axis, the resulting Zeeman term can be modeled as $H_\perp = \alpha_z B_z \sigma_z$. In this case, the dispersions become $E_\pm = \pm(v^2\hbar^2 k_y^2 + \alpha_z^2 B_z^2)^{1/2}$, i.e., the Dirac point is gapped. Likewise, when the magnetic field is applied along the $x$-axis, the Dirac spectrum is similarly gapped.

**Extended discussion on the band topology of Bi$_4$Br$_4$**

For a [001] monolayer, it is a standard 2D $Z_2$ topological insulator (TI) with a helical edge state protected by time-reversal symmetry[29]. Since this 2D band topology has been well discussed in the main text, below we elaborate the 3D band topology of bulk Bi$_4$Br$_4$. In short, there are two types of topological boundary states known for this time-reversal-invariant (i.e., non-magnetic) spin-orbit-coupled bulk material: (i) higher-order hinge states and (ii) surface Dirac cones.

(i) The higher-order band topology of 3D Bi$_4$Br$_4$ is protected by time-reversal and inversion symmetries. The resulting symmetry indicators for Bi$_4$Br$_4$ are $Z_{2,2,2,4} = \{0,0,0,2\}$[32]. This means that the material is neither a strong TI nor a weak TI. Rather, it features a double band inversion in the bulk and a single protected helical hinge state on an inversion-symmetric path on the surface of any inversion-symmetric crystal. Which hinges feature this state exactly depends on the surface termination and cannot be predicted from the bulk electronic structure and its topological invariants[34]. Moreover, the helical hinge state can be protected by time-reversal symmetry alone, indicating that the hinge states can survive under weak inversion symmetry breaking and in crystals with inversion asymmetric terminations[34].

(ii) 3D Bi$_4$Br$_4$ also has a nontrival $Z_2$ topological invariant protected by the C$_2$ rotation symmetry around the [010] axis, which is the atomic chain direction. This features a single band inversion in each of the two C$_2$ invariant subspaces[32]. It guarantees the existence of two gapless Dirac cones on any surface that preserves the C$_2$ rotation symmetry, i.e., the [010] and [0-10] surfaces. When the C$_2$ rotation symmetry is broken at these surfaces, the surface states become gapped[34].

Therefore, an inversion- and C$_2$- symmetric rod-like crystal with two capping surfaces normal to the [010] direction would feature both, helical hinge states along the [010] direction and gapless Dirac cones at the [010] and [0-10] surfaces, as illustrated by Extended Data Fig. 10. Such a geometry shows that the two topological invariants are not completely independent: if a capping surface was not gapless, there is no possibility to connect the hinge states in a way that preserves the C$_2$ symmetry, given that the hinge states need to form a singly connected loop due to the spectral flow they carry. Moreover, when the C$_2$ rotation symmetry is broken at the capping surfaces, the surface states become gapped, and their "residue" forms the gapless hinge states connecting the original hinge states, as illustrated by Extended Data Fig. 10.

In our experiment, we do not probe the [010] surface. In general, it would be challenging to probe this surface, as a crystal termination perpendicular to the atomic chains does not naturally form a clean surface, and because most likely the $C_2$ symmetry is unavoidably broken there. Thus, the [001] surface and the [100] surface (as part of the step edge) that we do probe do not carry any gapless surface states. This enables us to measure only the gapless hinge states in the bulk gap.

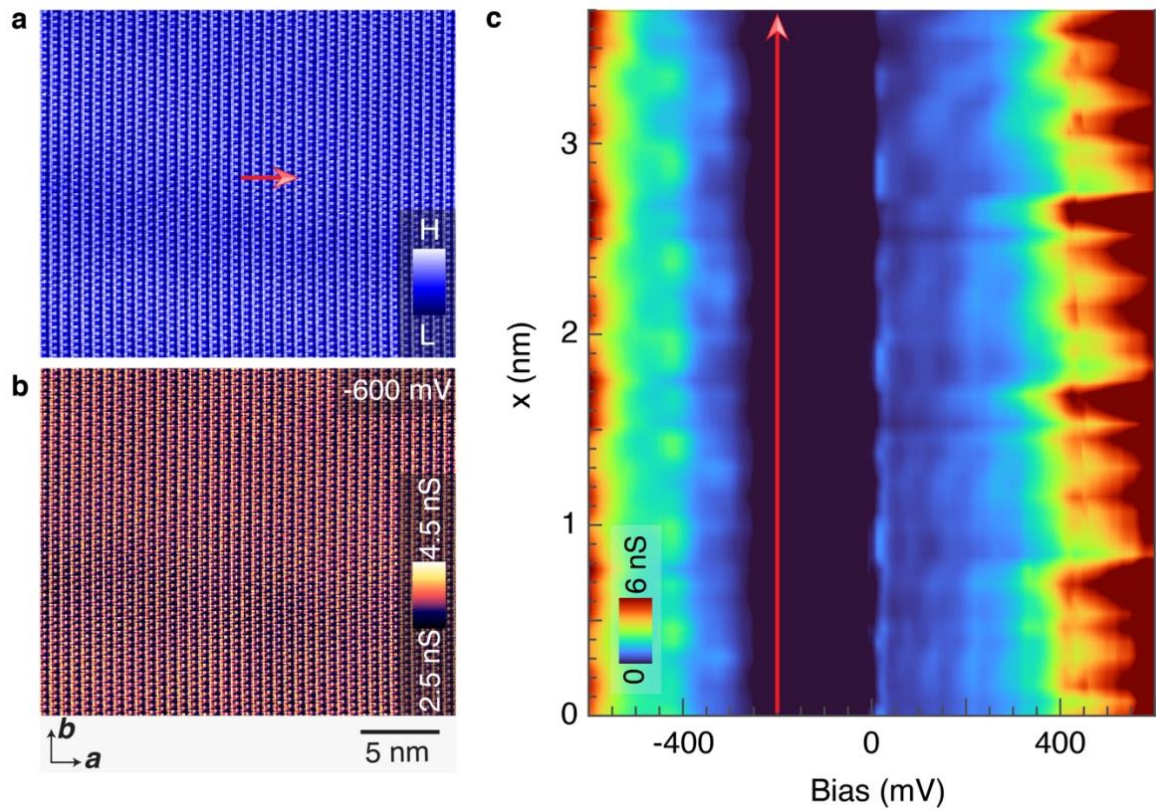

**Extended Data Figure 1. Intra-unit cell modulation of differential conductance. a**, Topography image of a clean region. **b,** Corresponding dI/dV map taken at V = 600mV. **c,** Intensity plot of the dI/dV spectrum taken along the line in **a.**

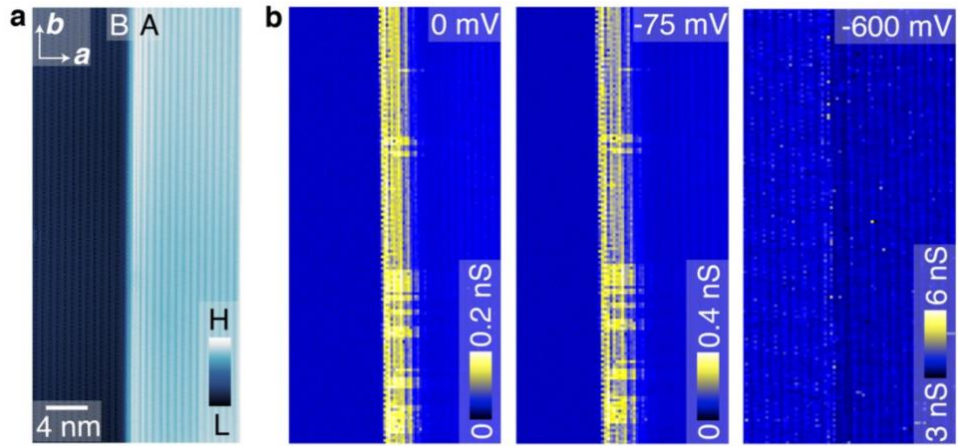

**Extended Data Figure 2. Bias dependence of the monolayer edge state. a**, Topographic image of an monolayer step edge. **b**, corresponding dI/dV maps, taken at V = 0mV, -75mV, and -600mV.

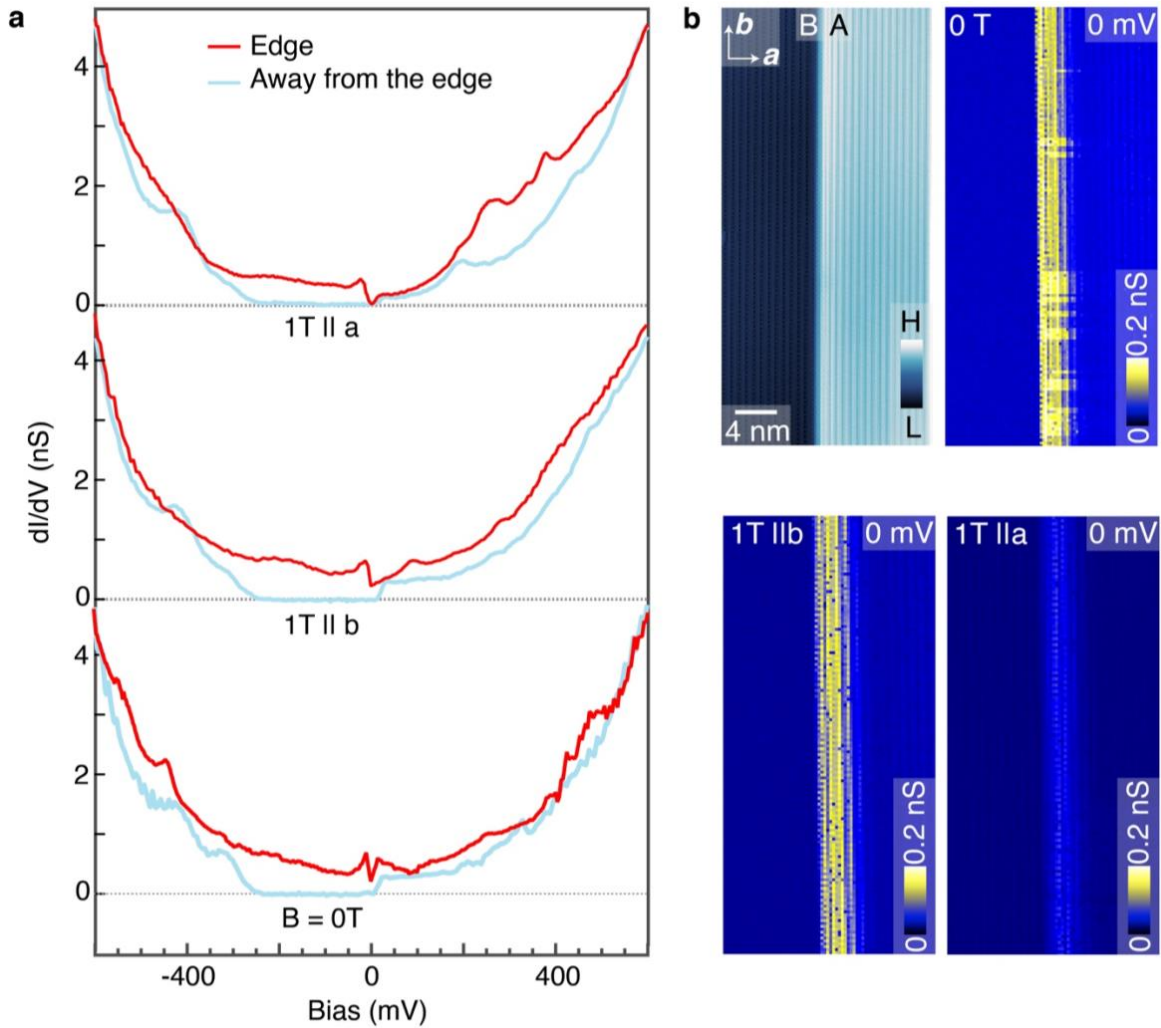

**Extended Data Figure 3. Anisotropic in-plane magnetic field response of the edge state. a**, Field dependent differential spectra taken on the edge and away from the edge, denoted with red and blue curves, respectively. Spectra are offset for clarity. **b**, Topography and the corresponding field dependent dI/dV maps (shown by using the same color scale) further elucidating the anisotropic in-plane magnetic field response of the edge state.

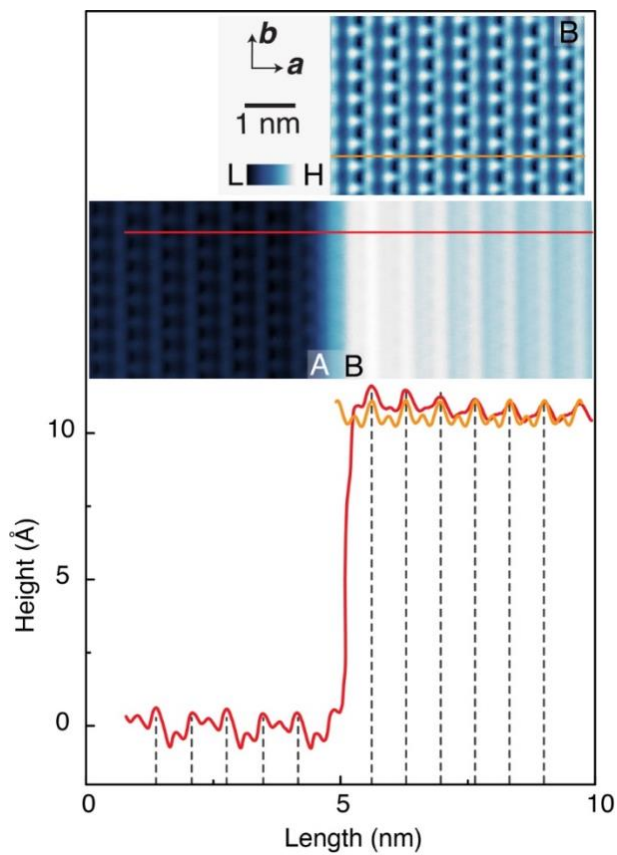

**Extended Data Figure 4. High-resolution imaging of the monolayer step edge.** We show the height profile across a monolayer step edge (marked on lower inset image), in comparison with that from a surface far from step edge (mark on the upper inset image).

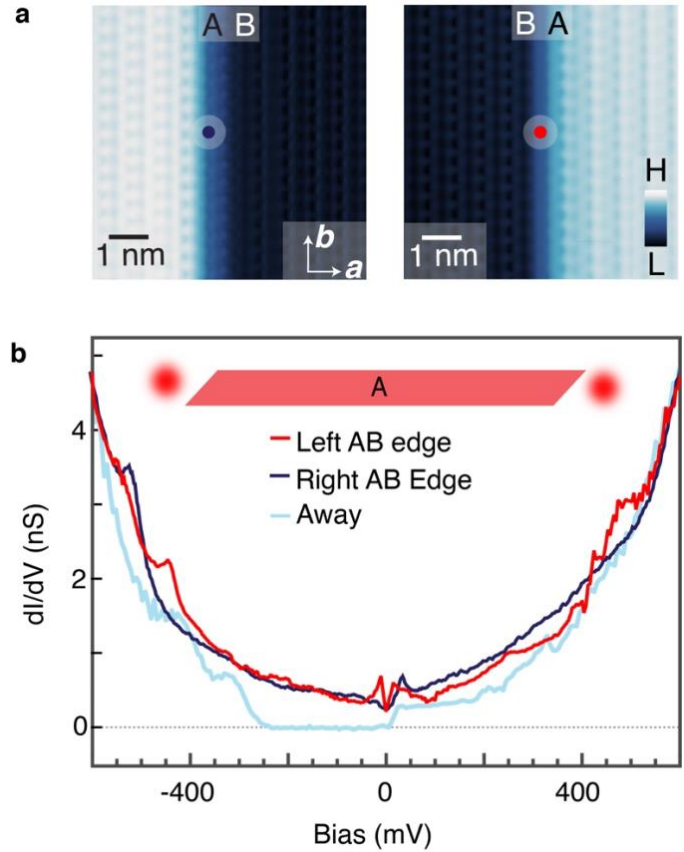

**Extended Data Figure 5. Tunnelling spectroscopy for both edges of a monolayer step**. **a**, Topographic image of right and left monolayer step edges (AB type). **b**, Differential spectra taken at the left AB monolayer step edge (red), at the right AB monolayer step edge (violet), and away from the edges (blue) reveal the presence of gapless edge states in both AB edges. The inset shows the schematic of the monolayer step edge states.

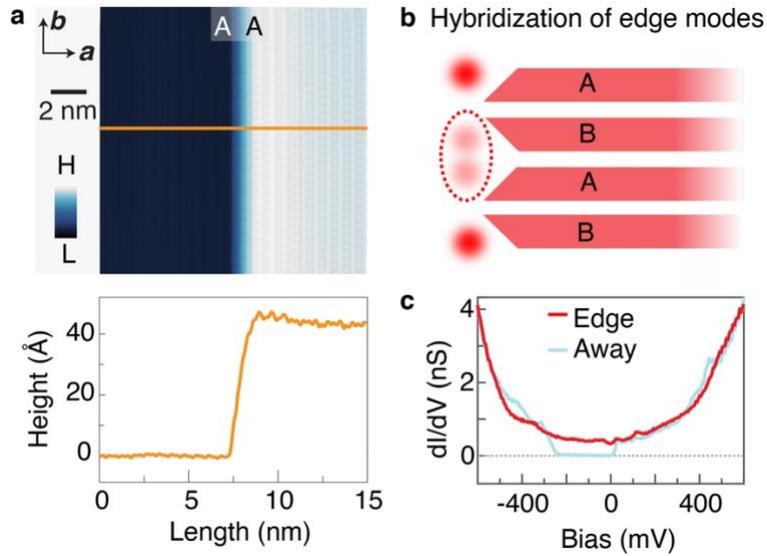

**Extended Data Figure 6. Hinge state candidate for a four-layer step edge. a**. Topographic image and corresponding height profile of a four-layer step edge. **b**, Schematic of a four-layer edge showing quantum hybridization of the quantum spin Hall edge states for neighboring layers. The destructive hybridization is illustrated by the lighter color of the edge states (red spheres). **c**, Differential spectra taken at the step edge (red) and away from the edge (blue) providing evidence for a candidate gapless hinge state.

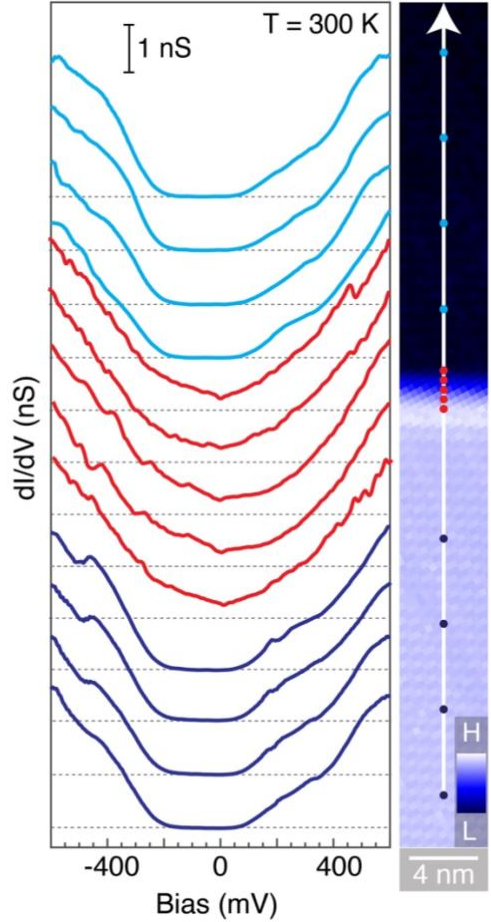

**Extended Data Figure 7. Spectroscopy of the monolayer edge state at room temperature.** Differential spectra of the edge (red curves) and surface (dark and light blue curves), taken along the *a*-axis direction (marked on the corresponding topographic image in the right panel with a white line), exhibit an insulating gap away from edge and gapless edge state at T =300 K. Spectra are offset for clarity, and their real-space locations are marked on the topography with color-coded dots. Dashed horizontal lines indicate zero density of states of the corresponding spectra.

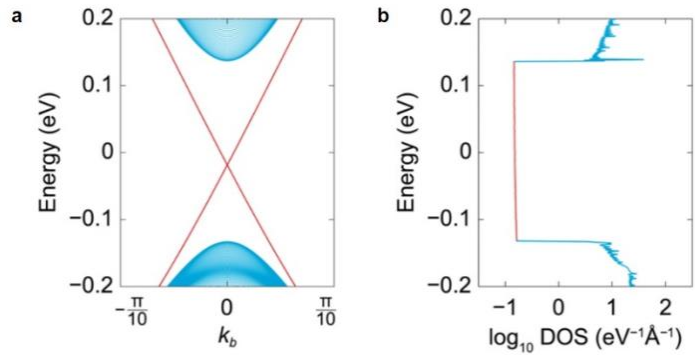

**Extended Data Figure 8. Calculated helical edge state at a one-layer step edge. a,** The edge-projected band structure for a (001) monolayer ribbon on the top surface of $Bi_4Br_4$. The cyan bands are from the bulk and (001) surfaces of the system. The red bands are the helical edge states. Due to the inversion symmetry

of the monolayer, the bands are doubly degenerate at each $k_b$. The ribbon is infinitely long in the *b* direction and 200-chain wide in the *a* direction. **b,** The DOS plotted in a log scale corresponding to the band structure in **a**.

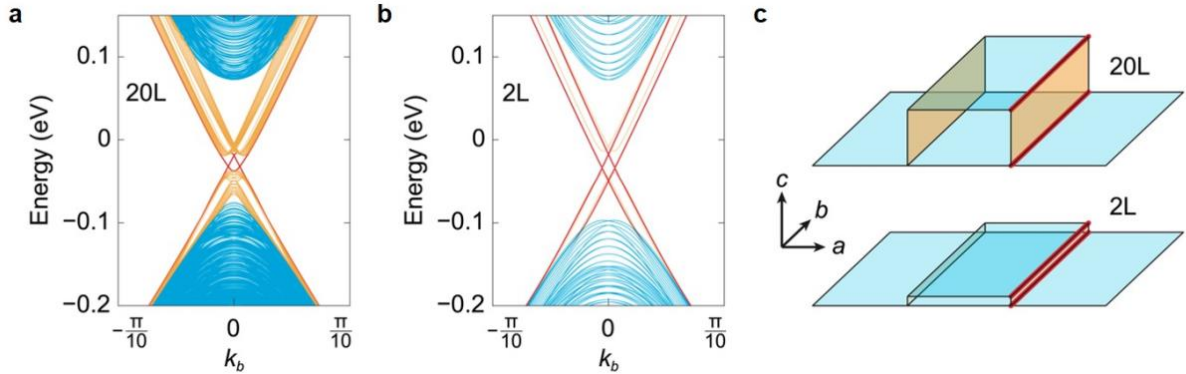

**Extended Data Figure 9. Calculated bilayer and twenty-layer step-edge states. a,** The edge-projected band structure for a (001) twenty-layer ribbon on the top surface of $Bi_4Br_4$. The cyan bands are from the bulk and (001) surfaces of the system. The orange bands are from the (100) and (-100) side surfaces of the ribbon. The red bands are the gapless hinge states. Due to the inversion asymmetry of even-layer systems, the bands are singly degenerate at each $k_b$. The ribbon is infinitely long in the *b* direction and 50-chain wide in the *a* direction. **b,** The same as a but for a bilayer ribbon. The two helical edge states from the two monolayers are gapped at one side (orange) but remain gapless at the other side (red). **c,** The real space schematics of the surface and hinge/edge states in **a** and **b**.

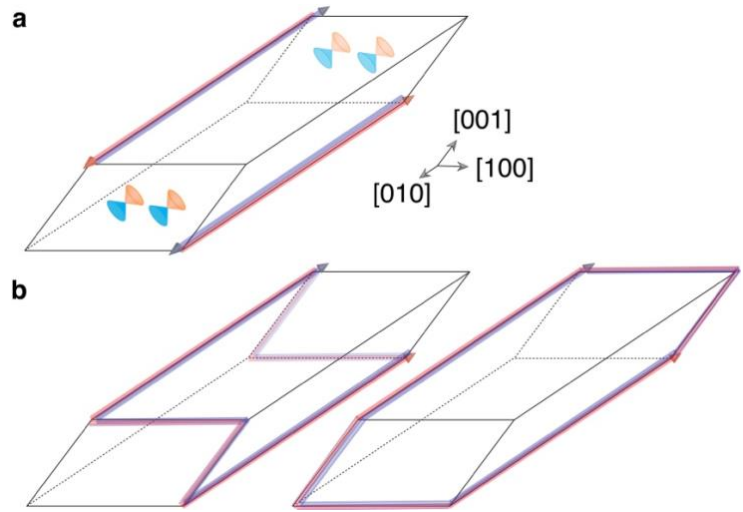

**Extended Data Figure 10. Topologically protected boundary states in a nanorod geometry with inversion symmetry and $C_2$ rotation symmetry around the [010] axis. a,** The nanorod of $Bi_4Br_4$ featuring helical hinge states from higher-order band topology and two surface Dirac cones on the [010] surface protected by the $C_2$ rotation symmetry. **b,** Two possible inversion symmetric paths for the helical hinge states when the $C_2$ symmetry is broken on the [010] surface.

**Acknowledgments:** Experimental and theoretical work at Princeton University was supported by the Gordon and Betty Moore Foundation (GBMF4547 and GBMF9461; M.Z.H.). C.Y. and F.Z. acknowledge the Texas Advanced Computing Center (TACC) for providing resources that have contributed to the research results reported in this work. The theoretical and computational work at UT Dallas was supported by the National Science Foundation under grant numbers DMR-1921581 through the DMREF program, DMR-1945351 through the CAREER program, and DMR-2105139 through the CMP program. F.Z. also acknowledges support from the Army Research Office under grant number W911NF-18-1-0416. T.N. acknowledges supports from the European Union's Horizon 2020 research and innovation programme (ERC-StG-Neupert-757867-PARATOP). T.A.C. was supported by the National Science Foundation Graduate Research Fellowship Program under grant no. DGE-1656466. Crystal growth is funded by the National Science Foundation of China (NSFC) (11734003), the National Key Research and Development Program of China (2016YFA0300600). Y.G.Y. is supported by NSFC (11574029) and the Strategic Priority Research Program of Chinese Academy of Sciences (XDB30000000).

**Author contributions:**
STM experiments were performed out by N.S., M.S.H, Y. J. M.L. in consultation with J.X.Y. and M.Z.H. Crystal growth was carried out by Z.W., Y.L., Y.Y., Z.Y., and S.J. TEM measurements were performed by G.C. and N.Y. Theoretical works were carried out by C.Y., F.Z., Y.C.L, T.R.C, T.N., H.L. in consultation with M.S.H, J.X.Y. and M.Z.H. Figure development and the writing of the paper were undertaken by N.S., M.S.H., J.X.Y. and M.Z.H. M.Z.H. supervised the project. All authors discussed the results, interpretation, and conclusion.

**Competing interests:** The authors declare no competing interests.

**Data and materials availability:** The data that support the findings of this study are available from the corresponding authors upon reasonable request.